# On a novel nanomaterial based on graphene and POSS: GRAPOSS

Luca Valentini[1], Marta Cardinali[1], Josè M. Kenny[1], Mirko Prato[2], Orietta Monticelli[3*]

[1]Dipartimento di Ingegneria Civile e Ambientale, Università di Perugia, Strada di Pentima 4, INSTM, UdR Perugia, 05100 Terni, Italy.

[2]Istituto Italiano di Tecnologia, Via Morego 30, 16163 Genova, Italy.

[3]Dipartimento di Chimica e Chimica Industriale, Università di Genova, Via Dodecaneso 31, 16146 Genova, Italy.

KEYWORDS. *Graphene, POSS, nanomaterial, surface functionalization*

**ABSTRACT:** A novel nanomaterial which consists of graphene sheets decorated with silsesquioxane molecoles has been developed. Indeed, aminopropyl-silsesquioxane (POSS-NH$_2$) has been employed to functionalize graphene oxide sheets (GOs). The surface grafting of GOs with POSS-NH$_2$ has been established by infrared spectroscopy and X-ray photoelectron spectroscopy, while the morphology has been investigated by field emission electron microscopy as well as by atomic force microscopy. The combination of the amino functionalized POSS molecules with GO sheets produces a hybrid silicon/graphite-based nanomaterial, named GRAPOSS, for which the electrical conductivity of reduced GO was restored, thus allowing promising exploitations in several fields such as polymer nanocomposites.

Over the past decade, nanomaterials, namely structures characterized by very small feature size in the range of 1-100 nanometers (nm), have arisen a relevant interest, as they have the potential for wide-ranging industrial, biomedical, and electronic applications.[1] Indeed, the development of novel nanomaterials, which can be metals, ceramics, polymeric or composite materials, represents a current key research topic. Among the most recently studied materials, graphene has emerged as a subject of enormous scientific interest due to its exceptional electron transport, mechanical and high surface area.[2-4] In order to produce graphene, graphite oxide, namely a layered material produced by the oxidation of graphite has been widely exploited. In contrast to pristine graphite, graphene oxide (GO) sheets are heavily oxygenated, bearing hydroxyl and epoxide functional groups on their basal planes, in addition to carbonyl and carboxyl groups located at the sheet edges.[5,6] The presence of these functional groups makes graphene oxide sheets strongly hydrophilic, which allows graphite oxide to readily swell and disperse in water. Previous studies have shown that a mild ultrasonic treatment of graphite oxide in water results in its exfoliation to form stable aqueous dispersions that consist almost entirely of 1-nm-thick sheets.[7] In fact, at present, exfoliation of graphite oxide is the only way to produce stable suspensions of quasi-two-dimensional carbon sheets, making this a strategic starting point for large-scale synthesis of graphene sheets. As such, graphite oxide has recently attracted attention as a filler for polymer nanocomposites.[8-10]

Concerning the new developments in the field of silicon-based nanomaterials, within the last 10-15 years, the silsesquioxane chemistry has grown dramatically. Among this class of materials, polyhedral oligomeric silsesquioxanes (POSS) have attracted a significant research effort.[11] POSS are organic/inorganic molecules, sizing approximately 1 to 3 nm, with general formula (RSiO$_{1.5}$)$_n$ where R is hydrogen or an organic group, such as alkyl, aryl or any of their derivatives.[12-14] Indeed, these cubic siloxane cages with functionalized substituents have thus become very popular as nanometer-scale building blocks in a wide range of polymeric materials.[15-17]

Clearly, nanostructured hybrids derived from the combination of carbon-based nanomaterials and POSS may potentially merge the properties of the two starting nanomaterials.

Although, the combination of POSS with carbon nanotubes was already studied, to the best of the authors' knowledge, no reports concerning the study of graphene/POSS hybrids has not been reported so far. Indeed, carbon nanotubes of different diameters (1-3 nm) were filled with cube-shaped octasilasesquioxane (H$_8$Si$_8$O$_{12}$) by heat-vacuum deposition or using supercritical carbon dioxide.[18] It was shown that the interaction of the POSS molecules with the nanotubes depends on the diameter of the latter. H$_8$Si$_8$O$_{12}$ may also be encapsulated in single- and double-walled nanotubes using solution and ultrasonic methods.[19] Another study involving carbon nanotubes was recently reported, where multiwalled carbon nanotubes having acid chloride groups were functionalized by the external grafting of aminopropylisooctyl-POSS via amide linkages.[20] The resulting materials were blended with poly(lactic acid) with a better dispersion in the matrix due to the presence of the POSS components.

In this work, for the first time, we report on a direct coupling of amino-POSS (POSS-NH$_2$) with graphene oxide sheets in a solvent, namely tetrahydrofuran (THF), capable of completely solubilizing silsesquioxane molecules and dispersing GO (see Supporting Information).

The chemical changes occurring upon treatment of POSS-NH$_2$ THF solution with GO can be observed by FTIR spectroscopy as both neat POSS-NH$_2$ and GO display characteristic IR spectra.

Figure 1a is the FTIR spectrum of POSS-NH$_2$. The wide and intensive peak at 3350 cm$^{-1}$ is attributed to the characteristic absorption of -NH$_2$ groups. The strong double peaks at 2930 and 2873 cm$^{-1}$ correspond to the C-H stretching of the CH$_2$ groups in the organic corner groups of the cage structure. The absorption band at 1120 cm$^{-1}$ is the characteristic vibration of Si-O-Si bond. The absorption peak at 1030 cm$^{-1}$ is attributed to the special characteristic vibration of silsesquioxane cage Si-O-Si framework.[21] The peak at 760 cm$^{-1}$ is the bending vibration of Si-C bond in Si-CH$_2$.[22]

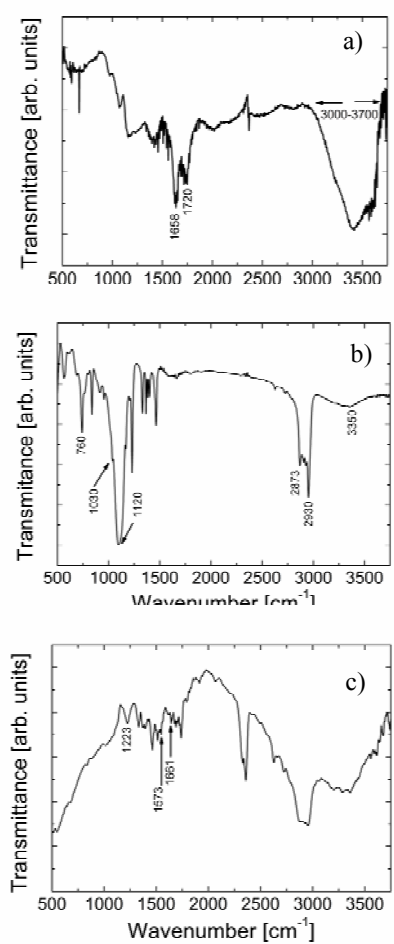

Figure 1. FTIR spectra of a) as-received POSS-NH$_2$, b) GO and c) GRAPOSS.

The most characteristic features in the FTIR spectrum of GO (Figure 1b) are the adsorption bands corresponding to the C=O carbonyl stretching at 1720 cm$^{-1}$, sp$^2$-hybridized C=C in plane stretching at 1550-1650 cm$^{-1}$. A broad and intense signal between 3000 and 3700 cm$^{-1}$ can be assigned to hydroxyls with contributions from COOH and H$_2$O (i. e. C-OH).[23]

The IR spectrum of GRAPOSS (Figure 1c), shows the disappearance of the band at 1720 cm$^{-1}$ and a corresponding appearance of a band with lower frequency (1661 cm$^{-1}$) assigned to the amide carbonyl stretch. In addition, the presence of new bands at 1573 and 1223 cm$^{-1}$, corresponding to N-H in-plane and C-N bond stretching, respectively, further confirms the presence of the amide functional group. Similar results for amine-functionalized carbon nanotubes have been previously reported in the literature.[24,25]

XPS can provide detailed information on the chemical composition and on the oxidation state of the surface species of neat POSS-NH$_2$, GO and of GRAPOSS. Wide scans, performed at a relatively low energy resolution, highlight the presence of Si species in GRAPOSS samples, suggesting that POSS-NH$_2$ functionalization of GO sheets occurred (see details and Figure S3 in the supporting information).

More details can be obtained by looking at higher resolution spectra collected on specific binding energy (BE) regions. In Figure 2, C 1s spectra of GO and GRAPOSS samples are compared.

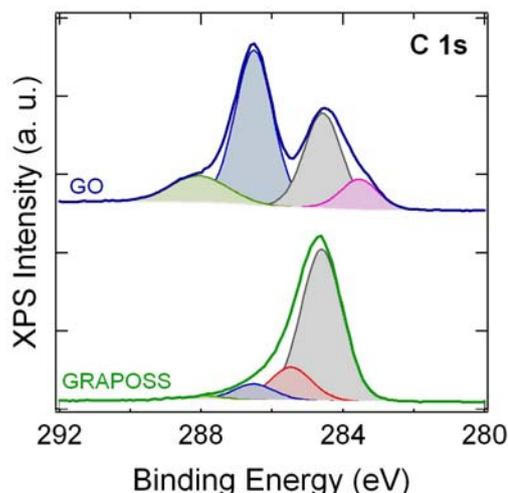

Figure 2. High resolution XPS spectra of GO and GRAPOSS: comparison of C 1s bands. Data are reported as continuous lines.

In the GO case, C 1s band is characterized by the presence of two clear maxima separated by approx. 2 eV, in agreement with results reported in literature.[26,27] The C 1s band can be decomposed in four components, at BE of 284.6, 286.6, 288.1 and 283.6 eV, which can be assigned respectively to graphitic carbon (C=C), carbon singly bound to oxygen (C-O-C, C-O-), carbonyl groups (C=O)[26,27] and to carbon vacancies.[28] Oxidized carbons represent approx. the 60% of the total amount of carbon species in GO.

The GRAPOSS C 1s spectrum appears completely different: the low BE component disappeared and the intensity of peaks related to oxidized species dramatically decreased (amount of oxidized carbons approx. 10% of carbon species in GRAPOSS) indicating an effective reduction of GO due to the POSS-NH$_2$ functionalization. In addition, a new component is present at 285.5 eV, corresponding to C-N bonds,[29] confirming once more the presence of amide functional groups.

The morphologies of the samples obtained from aqueous dispersions of GO and GRAPOSS were evaluated using field emission scanning electron microscopy. The prepared samples were also investigated by atomic force microscopy (see Supporting Information). What is expected is that the presence of carbonyl and carboxyl functional groups makes graphene oxide sheets strongly hydrophilic, which allows graphite oxide to swell and disperse in water leading to a re-stacking of the exfoliated graphene sheets due to the removal of the water. Accordingly what is observed for the sample obtained from the GO water solution was the formation of a extended area paper-like structure (Figure 3a) replaced by GRAPOSS structure consisting of isolated flakes (Figure 3b).

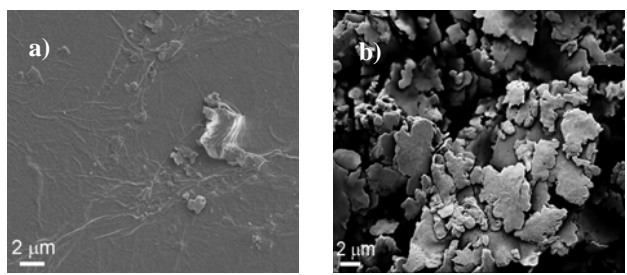

Figure 3. FE-SEM images of samples deposited from dispersions of a) GO in water and b) GRAPOSS.

GO is an electrically insulating material consisting of a large number of C-O bonds. Removal of C-O bonds by chemical reduction technique produces reduced graphene oxide sheets and allows one to restore the electrical properties.

Figure 4 summarizes the real part of the complex conductivity, $\sigma'$, at room temperature for GO, POSS-NH$_2$ and GRAPOSS samples, respectively.

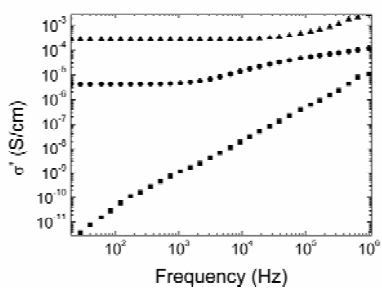

Figure 4. Real component of the complex conductivity as a function of frequency for the GO (■), POSS-NH$_2$ (●) and GRAPOSS (▲) samples at room temperature.

Pure POSS-NH$_2$ shows frequency dependent conductivity behavior over the entire frequency range indicating that these systems exhibit a dielectric or insulating behavior. A change in the frequency behavior of $\sigma'$ occurs for the pure GO and GRAPOSS samples; at low frequencies, $\sigma'$ is frequency independent up to a characteristic frequency, at which point $\sigma'$ increases with frequency being the crossover frequency shifted to higher frequencies for the GRAPOSS sample. This characteristic is reflective of a insulator-conductor transition due to the partial chemical reduction of GO sheets.

In conclusion, a novel carbon/silicon nanomaterial, based on graphene and POSS, has been developed. Indeed, the reaction between a silsesquioxane characterized by an amino group as reactive side and graphene oxide sheets was used to graft POSS onto GO sheets. We observed that GO sheets show a charge trap behavior that was partially passivated by the grafting of silsesquioxane molecules.

**Supporting Information**. Experimental procedures, AFM and XPS characterization of the prepared samples. This material is available free of charge via the Internet at http://pubs.acs.org.

## AUTHOR INFORMATION

**Corresponding Author**
* E-mail: orietta.monticelli@unige.it, Tel.: 0039-010-3536196.

**ACKNOWLEDGMENT**
We are grateful to the Italian Ministry of Education and University through the 2008 PRIN project (Grant No. 20089B75ML_001).

**REFERENCES**
(1) Li, B.; Zhong, W.-H.; Shatkin, J. A.; Dang, V. T.; Maguire, R. G., Nanoscience and Nanomaterials: Synthesis, Manufacturing and Industry Impacts 2011,….
(2) Allen, M. J.; Tung, V. C.; Kaner, R. B. *Chem. Rev.* **2010**, *110*, 132.
(3) Kim, H.; Abdala, A. A.; Macosko C. W. *Macromolecules* **2010**, *43*, 6515.
(4) Dreyer, D. R.; Park, S.; Bielawski, C. W.; Ruoff R. S. *Chem. Soc. Rev.* **2010**, *39*, 228.
(5) Stankovich, S.; Dikin, D. A. R. D.; Kohlhaas, K. A.; Kleinhammes, A.; Jia, Y.; Wu, Y.; Nguyen, S. T.; Ruoff, R. S. *Carbon* **2007**, *45*, 1558.
(6) Lerf, A.; He, H.; Forster, M.; Klinowski, J. *J. Phys. Chem. B* **1998**, *102*, 4477.
(7) Bourlinos, A. B.; Gournis, D.; Petridis, D.; Szabo, T.; Szeri, A.; Dekany, I. *Langmuir* **2003**, *19*, 6050.
(8) Cote, L. J.; Cruz-Silva, R.; Huang, J. *J. Am. Chem. Soc.* **2009**, *131*, 11027.
(9) Cai, D.; Yusoh, K.; Song, M. *Nanotechnology* **2009**, *20*, 085712.
(10) Ding, Q.; Liu, B.; Zhang, Q.; He, Q.; Hu, B.; Shen, J. *Polym. Int.* **2006**, *55*, 500.
(11) Cordes, D. B.; Lickiss, P. D.; Rataboul, F. *Chem. Rev.* **2010**, *110*, 2081 .
(12) Harrison, P. G. *J. Organomet. Chem.* **1997**, *542*, 141.
(13) Baney, R.H.; Itoh, M.; Sakakibara, A.; Suzuki, T. *Chem. Rev.* **1995**, *95*, 1409.
(14) Voronkov, M. G.; Lavrentyev, V. I. *Top. Curr. Chem.* **1982**, *102*, 199.
(15) Fina, A.; Monticelli, O.; Camino G. *J. Mater. Chem.* **2010**, *20*, 9297.
(16) Monticelli, O.; Fina, A.; Ullah A.; Waghmare, P. *Macromolecules,* **2009**, *42*, 6614.
(17) Monticelli, O.; Fina, A.; Cozza, E. S.; Prato, M.; Bruzzo, V. *J. Mater. Chem.* **2011**, *11*, 18049.
(18) Wang, J.; Kuimova, M. K.; Poliakoff, M.; Briggs, G. A. D.; Khlobystov, A. N. *Angew. Chem., Int. Ed.* **2006**, *45*, 5188.
(19) Liu, Z.; Joung, S.-K.; Okazaki, T.; Suenaga, K.; Hagiwara, Y.; Ohsuna, T.; Kuroda, K.; Iijima, S. *ACS Nano* **2009**, *3*, 1160.
(20) Chen, G.-X.; Shimizu, H. *Polymer* **2008**, *49*, 943.
(21) Feher, F. J.; Newman, D. A.; Walzer, J. F. *J. Am. Chem. Soc.* **1989**, *111*, 1741.
(22) Zhang, Z. P.; Liang, G. Z.; Lu, T. L. *J. Appl. Polym. Sci.* **2007**, *103*, 2608.
(23) Acik, M.; Lee, G.; Mattevi, C.; Pirkle, A.; Wallace, R. M.; Chhowalla, M.; Cho, K.; Chabal, Y. *J. Phys. Chem. C*, DOI: 10.1021/jp2052618.
(24) Lambert, J. B.; Shurvell, H. F.; Lightner, D. A.; Cook, R. G. Organic structural spectroscopy; Prentice Hall: Upper Saddle River, NJ, 1998.
(25) Mawhinney, D. B.; Naumenko, V.; Kuznetsova, A.; Yates, J. T.; Liu, J.; Smalley, R. E. *J. Am. Chem. Soc.* **2000**, *122*, 2383.
(26) Schniepp, H. C.; Li, J.-L.; McAllister, M. J.; Sai, H.; Herrera-Alonso, M.; Adamson, D. H.; Prud'homme, R. K.; Car, R.; Saville, D. A.; Aksay, I. A. *J. Phys. Chem. B* **2006**, *110*, 8535.
(27) Paredes, J. I.; Villar-Rodil, S.; Martínez-Alonso, A.; Tascón, J. M. D. *Langmuir* **2008**, *24,* 10560.
(28) Larciprete, R.; Fabris, S.; Sun, T.; Lacovig, P.; Baraldi, A.; Lizzit, S. *J. Am. Chem. Soc.* **2011**, *133*, 17315.

(29) Stankovich, S.; Dikin, D. A.; Piner, R. D.; Kohlhaas, K. A.; Kleinhammes, A.; Jia, Y.; Wu, Y.; Nguyen, S. T.; Ruoff, R. S. *Carbon* **2007**, *45*, 1558.

# On a novel nanomaterial based on graphene and POSS: GRAPOSS

Luca Valentini, Marta Cardinali, Josè M. Kenny, Mirko Prato, Orietta Monticelli[*]

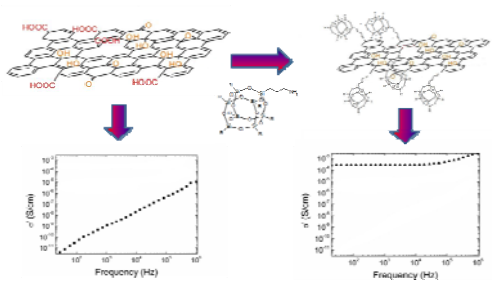